Representing Whole Slide Cancer Image Features with Hilbert Curves

Authors: Erich Bremer, Jonas Almeida, and Joel Saltz

Erich Bremer (corresponding author)
erich.bremer@stonybrook.edu
Stony Brook University
Biomedical Informatics

Jonas Almeida
jonas.dealmeida@nih.gov
National Cancer Institute
Division of Cancer Epidemiology & Genetics

Joel Saltz
Joel.Saltz@stonybrookmedicine.edu
Stony Brook University
Biomedical Informatics

**Abstract**
Regions of Interest (ROI) contain morphological features in pathology whole slide images (WSI) are delimited with polygons[1]. These polygons are often represented in either a textual notation (with the array of edges) or in a binary mask form. Textual notations have an advantage of human readability and portability, whereas, binary mask representations are more useful as the input and output of feature-extraction pipelines that employ deep learning methodologies. For any given whole slide image, more than a million cellular features can be segmented generating a corresponding number of polygons. The corpus of these segmentations for all processed whole slide images creates various challenges for filtering specific areas of data for use in interactive real-time and multi-scale displays and analysis. Simple range queries of image locations do not scale and, instead, spatial indexing schemes are required. In this paper we propose using Hilbert Curves simultaneously for spatial indexing and as a polygonal ROI representation. This is achieved by using a series of Hilbert Curves[2] creating an efficient and inherently spatially-indexed machine-usable form. The distinctive property of Hilbert curves that enables both mask and polygon delimitation of ROIs is that the elements of the vector extracted ro describe morphological features maintain their relative positions for different scales of the same image.

**Introduction**
Whole slide imaging (WSI) provides exquisite detail of entire microscope slides by scanning them at a high spatial resolution. Various programs have been developed to display these WSI creating a virtual microscope that can be overlaid with the original base WSI or additional overlays of extracted information, features, and annotations. In computer vision, image segmentation is the process of breaking down a digital image into multiple regions of interests (ROI) that are represented as circles, rectangles, or irregular polygon objects [1,3,4,6,7]. Employing machine vision techniques and algorithms such as deep learning neural networks and combined with relatively abundant processing power through modern CPU and GPUs, vast quantities of regions of interest in the form of polygons can be extracted for available WSI in a relatively short period of time. In the use case of nuclear segmentation, a half a million to a million and a half polygons can be extracted from each WSI. For each of these polygons, more than a dozen values can be derived such as perimeter, area, circumference, major and minor axis, etc.  Software packages like pyradiomics have been developed to further create feature characterization of the image pixels represented by the vast quantities of polygons.  Efficient storing, filtering, and retrieval of polygons create multiple challenges for software developers to create usable human interactions.

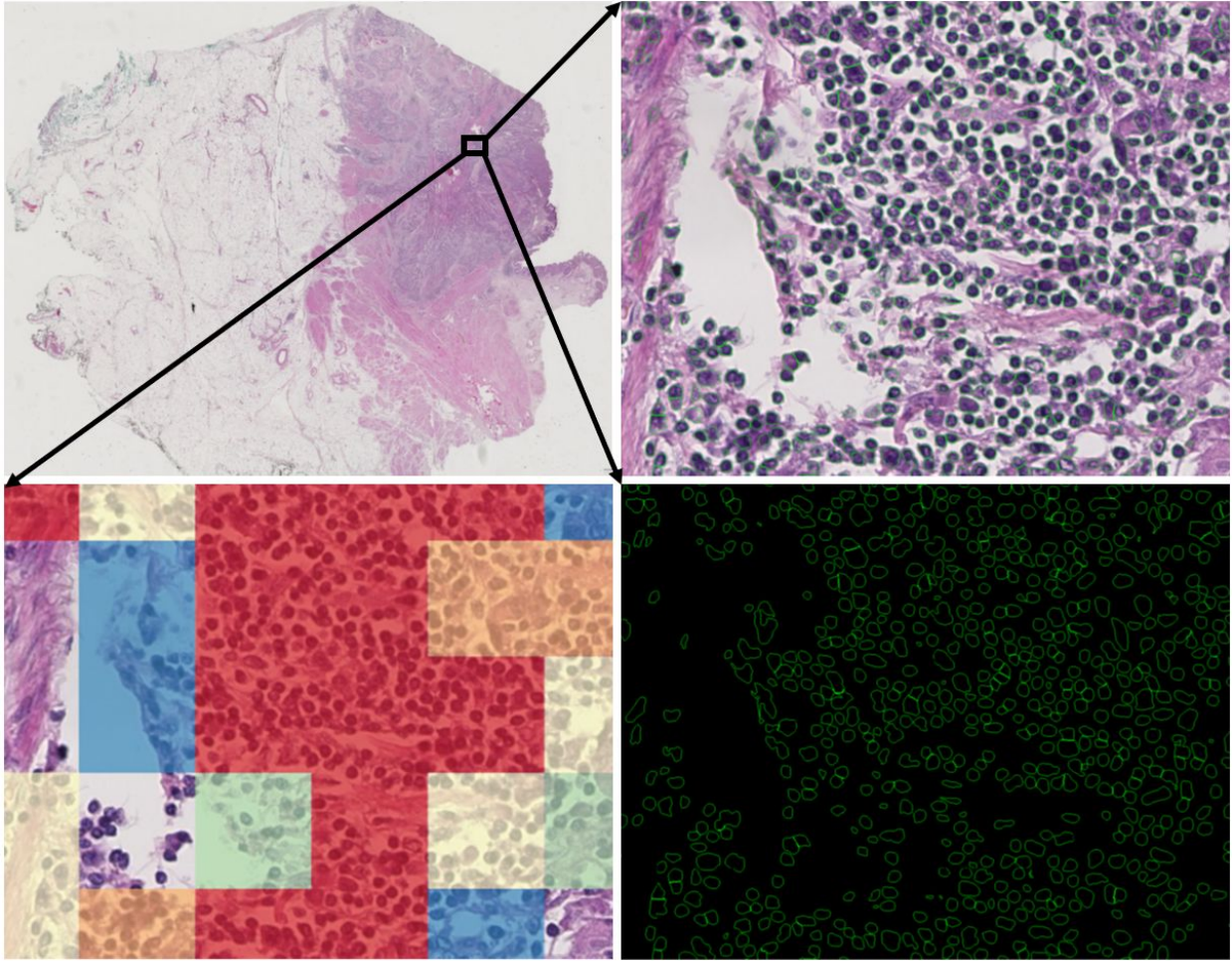

Figure 1 - ROI in a Cancer whole slide image (upper left).  Nuclear material ROI overlaid on the same image (upper right), nuclear material ROIs only (lower right), and a heatmap representation for tumor infiltrating lymphocytes (lower left).

**Technical Background**
Polygons are represented in various ways, usually, as sequentially connected points along the perimeter of the polygon the last point either implicitly connected to the first point in the list or the first point is explicitly listed at the end of the list.  The region inside this closed region would represent the region of interest.

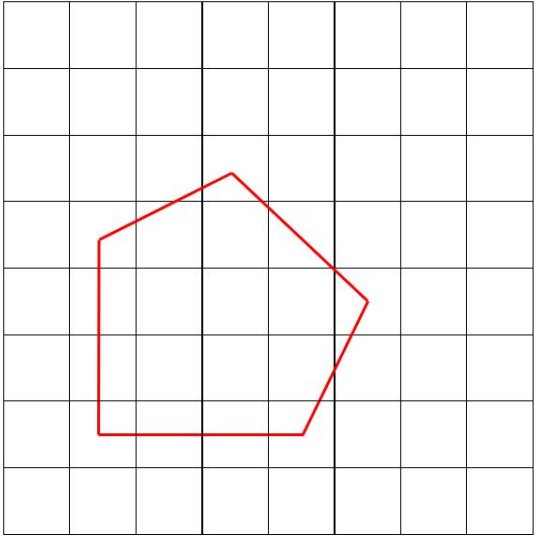

Figure 2 - simple irregular polygon

This is represented in various ways including but not limited the following textually notations:

JSON (JavaScript Object Notation)[5]
{"coordinates": [[[1,1],[1,4],[3,5],[5,3],[4,1],[1,1]]], "Type": "Polygon"}

Scalable Vector Graphics (SVG)[6]
<svg><polygon points="1,1 1,4 3,5 5,3 4,1 1,1" style="fill:lime;stroke:purple;stroke-width:1"/></svg>

Well-known text representation (WKT)[7]
POLYGON (1 1, 1 4, 3 5, 5 3, 4 1, 1 1)

An alternate method for ROI representation would be as a bitmap array with "zero" to represent background and a "one" to indicate a piece of the ROI.

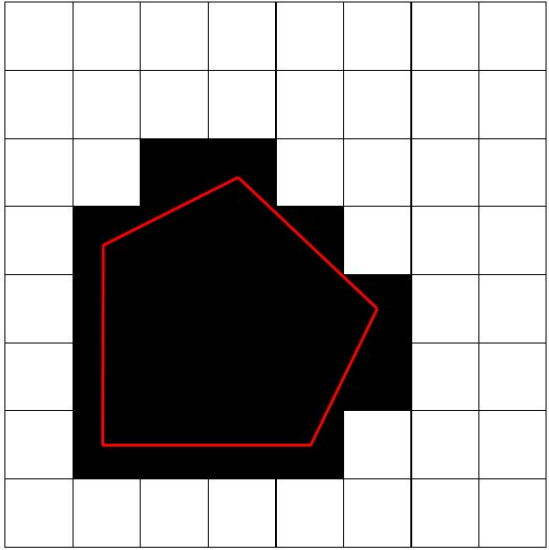

Figure 3 - bitmap representation of a polygon

At first glance, this would seem like a potential loss of resolution, however, due to requirements of various machine vision algorithms, it is often the bitmap that is the original data product with the textual representation being derived from the bitmap. Many software libraries exist to convert between these two representations as both polygon representations are often both employed.

As with most data, polygons are loaded into a database where spatial queries are done to subset that data as required. Not all databases employ advanced spatial indexing methods but all will support simple range queries for spatial object selection. For a particular rectangular search area S->($S_{minx}$, $S_{miny}$, $S_{maxx}$, $S_{maxy}$) with $S_{min}$ and $S_{max}$ being the upper left and lower right corners of the rectangular ROI, a SQL query would look as follows:

select polygon.id, polygon.x, polygon.y where
((polygon.x BETWEEN $S_{minx}$ AND $S_{maxx}$) and (polygon.y BETWEEN $S_{miny}$ AND $S_{maxy}$))

This approach works but performance erodes as billions of spatial objects in the database are approached depending on server CPU and IO horsepower. There are many spatial indexing algorithms such as geohash, hhcode, z-order curve, quadtree, octree, UB-tree, R-tree, R+ tree, R* tree, x-tree, kd-tree, m-tree, and bsp trees. Each has its advantages and disadvantages including algorithm coding complexity. For this paper, the Hilbert Curve was selected not only as a spatial indexing method but for actually representing the polygon itself rather than a textual form or as a binary mask.

A Hilbert curve is a continuous fractal space-filling curve that was first described in 1891 by German mathematician David Hilbert. In any rectangular grid or array, a Hilbert curve will visit each and every cell once and only once without the curve crossing over on itself. Hilbert curves give a mapping between 2D and 1D space that preserves a degree of locality[8].

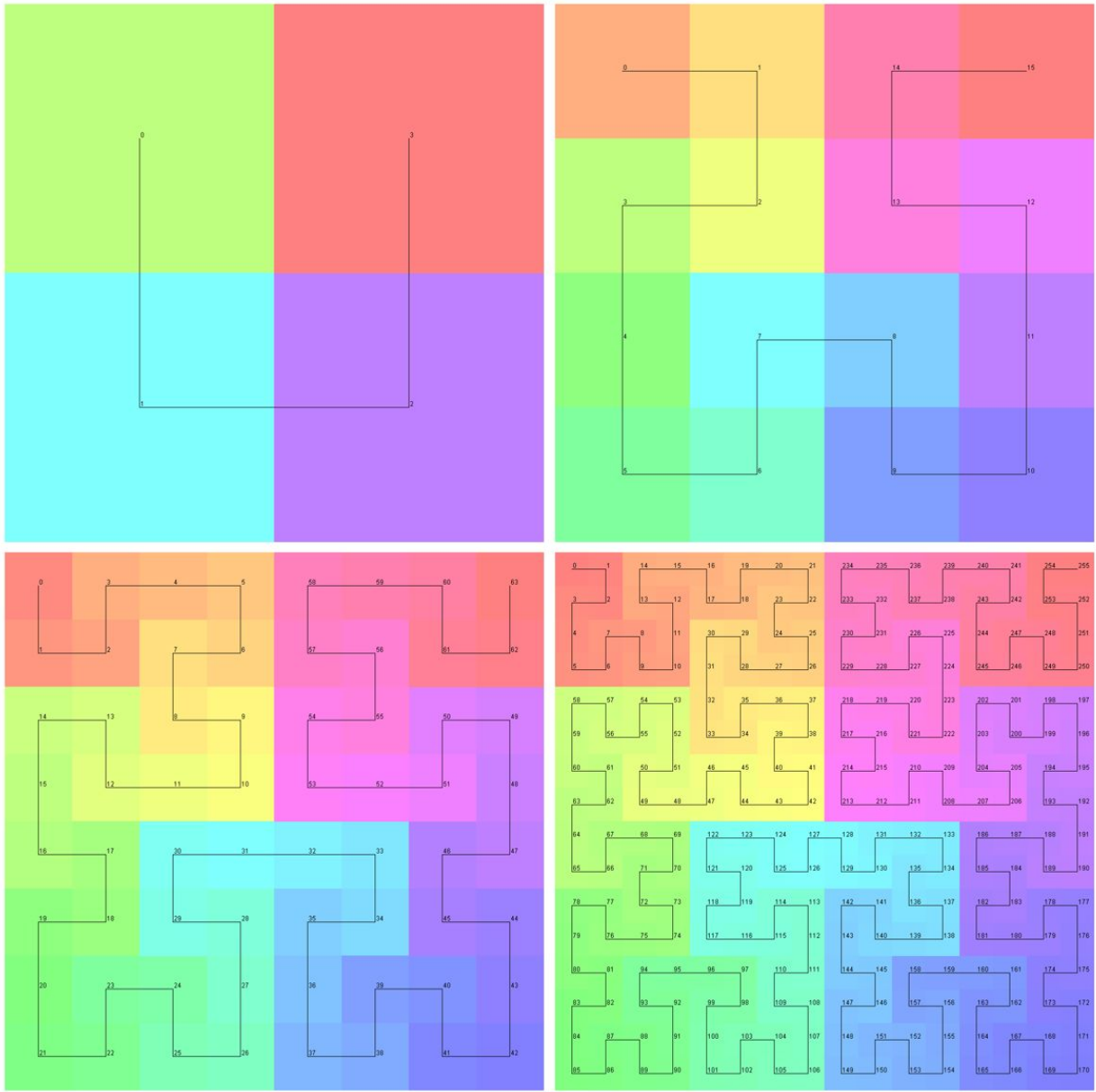

Figure 4 - First, second, third, and fourth order Hilbert Curves

**Approach**

The details of the mathematics of Hilbert curves is simplified for purposes of conceptual understanding as follows:
N = H(X,Y)
(X,Y) = H$^{-1}$(N)
Each and every 2D (X,Y) coordinate on a grid will map to a unique one dimensional value N. This process has an inverse function in that N can be mapped back to its original (X,Y) pair.

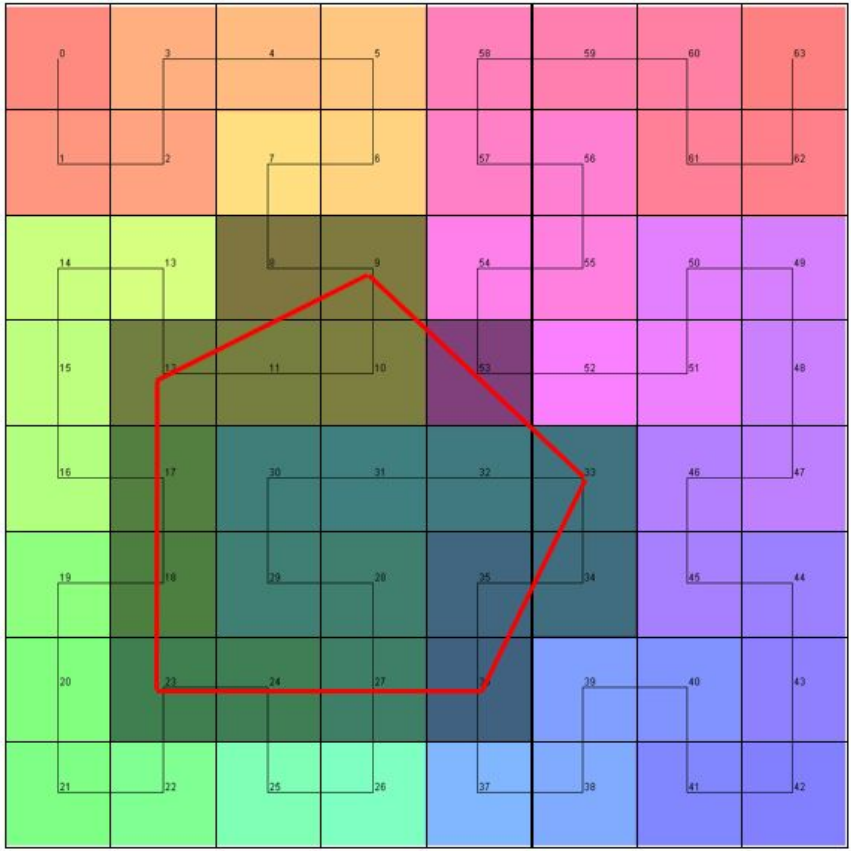

Figure 5 - intersected portions of Hilbert curve by ROI

Hilbert Polygon Representation in JSON for the above polygon (Figure 5)
{"name": "Polygon 1",
 "type": "Nuclear Material",
 "Ranges":[[8,12],[17,18],[23,24],[27,36],[53,53]]}

This representation of the polygon can be converted back by simply looking at the perimeter of the enclosed region. This is derived from the boundary algorithm for finding ranges from Moten, "Given an n-dimensional search region the exact hilbert curve ranges that cover the search region can be determined just by looking at the hilbert curve values on the perimeter (boundary) of the region."[9] For our purposes, we store our polygons solely in Hilbert form, search with Hilbert ranges, and only when the cartesian representation is needed, we use this algorithm to convert our Hilbert representations back to their cartesian versions.

Casting the two-dimensional points onto the one dimensional Hilbert curve has the advantage in that it can be layered into any database. Not all databases support spatial indexing but all support one-dimensional numeric indexes. Using this strategy, we can map each Hilbert Polygon range into a flat table that contains the beginning and end for each range and the ID of that range which can be mapped back to another table which contains the ID of the Hilbert Polygon which that range is a part of. The table is sorted on the field for the value of the start and end Hilbert ranges.

Take our original rectangular search area, S->($S_{minx}$, $S_{miny}$, $S_{maxx}$, $S_{maxy}$). This search area itself is a polygon and can also be converted to a Hilbert Polygon $S_{hp}$->($S_{1[b1,e1]}$,$S_{2[b2,e2]}$,$S_{3[b3,e3]}$,$S_{4[b4,e4]}$,$S_{5[b5,e5]}$,…) containing one or more being/end ranges. To obtain the list of polygons in $S_{hp}$, we simply take the union of the following:

select polygon.id, polygon.x, polygon.y where
    (polygon.hilbertvalue BETWEEN $S_{b1}$ AND $S_{e1}$) UNION

select polygon.id, polygon.x, polygon.y where
    (polygon.hilbertvalue BETWEEN $S_{b2}$ AND $S_{e2}$) UNION ...

select polygon.id, polygon.x, polygon.y where
    (polygon.hilbertvalue BETWEEN $S_{bn}$ AND $S_{en}$) UNION

**Results**

The Hilbert Curve polygon representation method was applied to an WSI image from The Cancer Genome Alas (TCGA) collection for the use case of nuclear segmentation. The irregular polygons extracted from the segmentation progress creates the most challenge to this methodology. Here below is a comparison of the two methods:

*Standard connected perimeter point polygons method*
The dimensions of the TCGA image were: 135,168 x 105,472 pixels
# of polygons extracted for nuclear segmentation = 1,547,170
# of points required to represent standard polygons = 54,600,980
35.3 (X,Y) points per polygon

*Hilbert representation method*
# of Hilbert ranges = 36,478,264
23.6 (N) Hilbert Ranges per polygon

**Discussion**
These numbers show a very promising potential use of the Hilbert Curve representation for ROI polygons, however, further analysis is required. Storage requirements are also a factor in a large-scale collection of ROI polygons. Most computer languages work with particular atomic number types for example take integer numbers - bytes, shorts, integers, longs. The differences between these types are in the number of bits used to represent the number which determines the range of values that can be represented. Further, in some systems and languages, the numbers can be limited to only signed numbers which could further limit the number of values for that particular data type since the leftmost bit is used for the sign. For instance, java "int" uses 4 bytes or 32 bits minus the sign bit which allows whole numbers from -2,147,483,648 to 2,147,483,647.

Nuclear segmentation polygons is the worst-case scenario for this methodology as more Hilbert ranges are needed to represent the finer details of complex irregular polygons. For our heatmap representations where our ROIs are all squares, far fewer Hilbert ranges are needed. If square sizes are selected as a power of 2, the number of ranges needed to represent that ROI drops to one potentially saving on storage requirements. The traditional representation of the same square would use 4 x (x,y) points but could be optimized for the rectangle/square by only needing the upper left and lower right points reducing the need to 2 x (x,y) points.  For an image of the above size, it would take at least 18 bits to represent one of the dimensions of the points # bits = ceil(ln(max(135168,105472))/ln(2)) = 18.  This would require 4-byte ints as opposed to 2-byte shorts.  Hilbert indexes essential inter-leave X,Y values into a one dimensional index which effectively doubles the storage requirements making them roughly equivalent in storage compared to a 2 x (x,y) rectangle but twice as good as expressing all four points.  The advantage of using the Hilbert Polygons for the square ROI case is that the representation is spatially indexed whereas the 2 x (x,y) rectangle would need further processing.

Further, Hilbert curves can not only represent polygons, but they can represent other annotation objects like points, lines, and lower-resolution representations of ROI polygons.

**Conclusion**
Using a sufficient number of Hilbert Ranges to express an irregular polygon allows a full-fidelity representation of the polygon without loss of resolution.  Coding requirements for implementing a Hilbert range search are fairly simple to code compared to other spatial indexing algorithms.  In various cases, storage requirements can be reduced while still providing spatially indexed data at the same time.  Hilbert Curves can also be generalized beyond 2 dimensions to n dimensional data.  Traditional textual and bitmap representations can be extracted from the Hilbert Polygons without loss of resolution.  In a follow up paper, we will demonstrate and discuss a software system that implements this Hilbert methodology for multiple types of ROI segmentations and classifications at scale.

**Financial support and sponsorship**
Nil.

**Conflicts of Interest**
There are no conflicts of interest.